\documentclass[prd,floatfix,twocolumn,amsmath,amssymb,nofootinbib]{revtex4}
\usepackage{}
\usepackage{graphicx,color,dcolumn,booktabs,bm}
\usepackage{longtable,lscape}
\usepackage{txfonts}
\usepackage{overpic}
\usepackage{amssymb}
\usepackage{epstopdf}
\usepackage{indentfirst}
\usepackage{feynmf}   
\usepackage{slashed}  
\usepackage{cases}
\usepackage{color}
\usepackage{float}
\usepackage{multirow}
\usepackage{graphicx,color,dcolumn,booktabs,bm}
\usepackage{epsfig,dsfont,amssymb,amsmath,amsfonts,amsbsy,mathrsfs}

\graphicspath{{Figures/}} %
\usepackage[colorlinks, citecolor=blue,anchorcolor=red,menucolor=red, linkcolor=red,filecolor=red,runcolor=red,urlcolor=blue,frenchlinks=red]{hyperref}

\begin{document}

\title{Regge-like spectra of excited singly heavy mesons}

\author{Duojie Jia}\email{jiadj@nwnu.edu.cn}
\author{Wen-Chao Dong}

\affiliation{Institute of Theoretical Physics, College of Physics and Electronic Engineering,
Northwest Normal University, Lanzhou 730070, China}
\begin{abstract}

In this work, we study the Regge-like spectra of excited singly heavy mesons by
proposing a general Regge-like mass relations in which the slope ratio $%
\alpha ^{\prime }/\beta ^{\prime }$ between the radial and angular-momentum
Regge trajectories is $\pi /2$ and the hadron mass undergoes a shift
including the heavy quark mass and an extra binding energy between heavy
quark and strange anti-quark. The relation is successfully tested
against the observed spin-averaged data of the singly heavy mesons in their
radially and angularly excited states. Some new predictions are made for
more excited excitations and the discussion is given associated with the QCD
string (flux tube) picture.

\end{abstract}


\maketitle

\section{Introduction}\label{sec1}

During the last two decades, heavy meson spectroscopy has
been greatly enlarged due to the discovery of numerous excited charm and
charm-strange states (charm mesons hereafter) by the B-Factory experiments
BaBar, Belle, CLEO and recently by the LHCb experiment \cite%
{Tanabashi:D18}. One of recent examples is the observations of the
heavy mesons $B(5970)$ by the CDF Collaboration in the $B^{+}\pi^{-}$
and $B^{0}\pi^{+}$ mass distributions \cite{Aaltonen:14} and the $B_{J}(5960)$
by the LHCb Collaboration in the $B\pi $ mass distributions \cite{AajiLHCb:2015}.
The first observation determined the mass of the $B(5970)$ resonances to
be $5978\pm 5\pm 12$ MeV for the neutral state and to be $5961\pm 5\pm 12$ MeV
for the charged state. The second observation, giving the mass of
$5969.2 \pm 2.9 \pm 5.3 $ MeV for the neutral state and $82.3 \pm 7.7 \pm 9:4$ MeV
for the charged state for $B_{J}(5960)$, seems to be consistent with the $B(5970)$
in their properties, and they may be the same state. Another example is the observation
of the charmed states of $D_{J}(3000)$ by the LHCb Collaboration in the $D\pi$
mass distribution data from pp collision \cite{Aaij:jhep13,Aaij:16}.
These experimental findings arouses theoretical interests to explore and
accommodate the newly-discovered states with various approaches, e.g., with
the potential quark models \cite%
{GodfreyM:D,KZLi:C15,SCLiu:C15,SCLiu:D15,SLM:D13,Wang:D11,Wang:D13,BeverenR:D10,BadalianB:D11,CloseTL:B07,GodfreyJ:D14,GodfreyM:D14,ChenYZ:D11}%
, Lattice QCD \cite{MoirPR:jp13,LewisW:D00} and other approaches (see Ref.
\cite{ChenCL:Rpp17} for a recent review). For the $B(5970)$, the two main
interpretations exist, with the assignments $2^{3}S_{1}$ \cite{SunSCL:D14}
and the candidate of $1^{3}D_{3}$ state \cite{XiaoZ:D14}. For the $%
D_{J}(3000)$ and $D_{J}^{\ast }(3000)$, there are the different
interpretations, the $D(3^{1}S_{0})$ and the $D(3^{3}S_{1})$ states,
respectively, or the candidates of $D(1F)$ or $D(2P)$. For the $D_{2}^{\ast
}(3000)$ the prediction favoring $D(3^{3}P_{2})$ was given \cite{WangCSL:D16}%
.

In Ref. \cite{KChenL:C18}, a comprehensive analysis was made for the whole
singly heavy systems of mesons and baryons using the Regge-like mass
relation,
\begin{equation}
(M-M_{Q})^{2}=\alpha L+c,  \label{HRegg}
\end{equation}%
with $M_{Q}$ the heavy quark mass, supporting that all heavy-light hadrons
fall on straight lines in a "shifted" orbital trajectory plane: $%
M\rightarrow M-M_{Q}$, with the slope ($\alpha $) nearly half of that for
the light hadrons. Here, $L$ is the orbital angular momentum of the hadron
system. This reflects that the notion of Regge trajectory, which is
widely-used in light-flavor sector and stems from the well-known Regge
theory of the scattering matrix \cite{Regge:59}, remains to be useful in the
spectra of the heavy flavor hadrons if one shifts the hadron mass $M$ by a
constant. In addition, the knowledge of these trajectories is also valuable
in the modeling of the recombination and fragmentation for hadrons
transition in the scattering region ($t<0$) \cite{Collons:77}. For the light mesons,
an updated picture emerges in Ref. \cite{Baker:D02} and in Refs. \cite%
{AllenGO:D01,Afonin:A07,Afonin:C07,Bicudo:D07,Anisovich:D00,Forkel:07a,Forkel:07b,Masjuan:D12}
where a more general linear Regge trajectories,
\begin{equation}
{\large M}^{2}{\large =\alpha J+\beta n+a}_{0},  \label{MnJ}
\end{equation}%
were proposed, with $n$ the principle quantum number and $a_{0}$ a constant.
This raises a question as to if the relation (\ref{MnJ}) applies for the
heavy mesons and how much is the slope ratio $\beta /\alpha $ is if it
applies. Differing from an existing pattern of approximate degeneracy ($%
\alpha \simeq \beta $) in the excited light hadron spectra \cite%
{Bugg:04,Afonin:A07,Afonin:C07,Anisovich:D00}, in the case of the heavy hadrons
with limited number of observed states, this type of degeneracy remains to be
explored, and study of possible pattern in them will be of interest for
accommodating or searching for some of the higher excitations of the heavy
mesons that have surfaced after the $B(5970)^{+/0}$ observation.

Purpose of this work is to propose and support that the relation (\ref{MnJ})
applies for the heavy-light (HL) mesons if the slope ratio $\beta /\alpha $
between the radial and angular-momentum Regge trajectories is chosen to be
$\pi /2$ and the mass $M$ of meson is shifted by a mount $M_{Q}-\tilde{\mu}%
_{lQ}$ with $M_{Q}$ the heavy quark mass and $\tilde{\mu}_{lQ}$ the
effective reduced mass of the heavy quark and light (anti)quark. We use the
mass relation to plot the orbital and angular-momentum Regge trajectories
combined for the observed spectroscopy of heavy-light mesons \cite%
{ChodosT:N74} and infer thereby that $D(3000)^{0}$,
$B_{J}(5840)$ and $B_{J}(5970)$ are most likely,
the $D(3P)$ and $B(2S)$ respectively. A semiclassical illustration for
the mass relation including the slope ratio $\pi /2$ is presented within the
picture of QCD string or flux tube. The flavor-dependence of the trajectory
parameters in the mass relation is noted.

It is well known that the parent linear Regge trajectories (\ref{MnJ}) with $%
n=0$ can be explained in the rotating string picture \cite%
{GoddardRB:B73,Nambu:D74,Goto:71}, where the quark and antiquark in meson
are assumed to be massless and tied by the gluon flux tube (QCD string). In
the massive case, the corrections to linear Regge trajectories were explored
in Refs. \cite{JohnsonN:PRD79,SelemW06} and recently in Refs. \cite%
{SonnenscheinW:JHEP14,BChen:A15}. When the radially excitations involved,
the issue of how well the Regge trajectory applies is of interest for understanding
the excited heavy mesons. For more discussions of the Regge-like relations
for the mesons, see \cite%
{AllenGO:D01,Afonin:A07,Afonin:C07,Bicudo:D07,Olsson:D97,Forkel:07a} and \cite%
{EFG:C10,Baker:D02} for instance.

In Section \ref{sec2}, we present the Regge-like mass relation for the heavy-light
mesons and outline the existing proposals that is directly related to the
our mass relation. In Section \ref{sec3}, we confront the mass relation proposed with
the observed masses of the excited HL mesons and thereby some of predictions
are given. A simple semiclassical picture that supports Regge-like mass
relation is outlined in the QCD string model in Section \ref{sec4} and the
conclusions are summarized in Section \ref{sec5}.

\section{Regge-like mass relation for orbitally and radially excitations}\label{sec2}

Our proposal for heavy-light (HL) mesons spectrum is to extend the
Regge-like relation (\ref{MnJ}) to the general case that applies for the
orbital and angularly excited states, by choosing the slope ratio between
the radial and angular-momentum Regge trajectories to be $\pi :2$, that is,
$\beta /a=$ $\pi /2$, and the mass $M$ of mesons to be shifted by $M_{Q}-%
\tilde{\mu}_{lQ}$ with $\tilde{\mu}_{lQ}=$ $km_{q}/(km_{q}+M_{Q})$ the
effective reduced mass of the heavy quark and light (anti)quark:
\begin{equation}
M\rightarrow M-M_{Q}+\tilde{\mu}_{lQ}=M-M_{Q}/(1+km_{q}/M_{Q}).
\label{shift}
\end{equation}%

Here, the prefactor $k$ is introduced to describe effectively the
scale-dependent mass running of the light quarks that may be relevant in the
higher excitations. The emergence of $\pi /2$ seems to be quite unusual in
the light of pure phenomenology. From the viewpoint of this work, however,
it stems from the fact that the energy cost for orbital motion of a string
linked to quarks differs itself from that for radial motion (see Section
\ref{sec4} for details). The main proposal is then
\begin{equation}
\left( M-\frac{M_{Q}}{1+km_{l}/M_{Q}}\right) ^{2}=\pi b\left( L+\frac{\pi }{2%
}n\right) +\left( m_{l}+\frac{P_{Q}^{2}}{M_{Q}}\right) ^{2},  \label{GRegg}
\end{equation}%
where $L$ is the quantum numbers of the orbital angular momentum, $n$ the
radial quantum number, $b$ the slope parameter, and $M$, $M_{Q}$ and $m_{l}$
are the masses of the HL meson, heavy quark and light quark, respectively.
In the intercept part of the relation (\ref{GRegg}),
\begin{equation}
P_{Q}\equiv M_{Q}v_{Q}\simeq M_{Q}\left( 1-\frac{m_{bareQ}^{2}}{M_{Q}^{2}}%
\right) ^{1/2},  \label{PQ}
\end{equation}%
that is,
\begin{equation*}
\text{Intercept}=\left[ m_{l}+M_{Q}\left( 1-\frac{m_{bareQ}^{2}}{M_{Q}^{2}}%
\right) \right] ^{2},
\end{equation*}%
where $m_{bareQ}=1.275GeV$ and $4.18$GeV are the bare masses of the heavy
quark $Q=c,b\,$, respectively.

It is useful to view Eq. (\ref{GRegg}) as an extension of the mass relation
suggested by Selem and Wilczek \cite{SelemW06} for the singly heavy hadrons,

\begin{equation}
M-M_{Q}=\sqrt{\frac{\alpha }{2}L}+2^{1/4}\kappa \frac{\mu ^{3/2}}{L^{1/4}},
\label{SW}
\end{equation}%
which implies $(M-M_{Q})^{2}\propto (\alpha /2)^{2}L$ when the orbital
angular momentum of hadrons $L\rightarrow \infty $ or the mass of light
quark $\mu \rightarrow 0$, as examined in Ref. \cite{KChenL:C18}. It agrees
with the relation (\ref{MnJ}) with $n=0$ up to a mass shift $M\rightarrow
M-M_{Q}\ $or $M\rightarrow M-\mathcal{O}(M_{Q})$.

In the light of Ref. \cite{SelemW06}, the established states of mesons and
baryons can be accommodated with appropriate quantum numbers and approximate
mass using the hypotheses of the loaded flux-tube (string) and the emergent
diquark. In the case of the HL systems, the hypotheses predict (\ref{SW})
for the hadron mass relation with $L$ (the orbital angular momentum) and $%
\mu $ (the effective mass of the light quark or diquark). The relation (\ref{SW}%
) works well for the excited D/B mesons, as examined in Ref. \cite%
{KChenL:C18}, but evidently, suffers from the singularity ($L=0$, the S-wave
states) occurred in the second term in its right hand side (RHS). One way to
avoid the singularity may be to view the $\kappa $-term in Eq. (\ref{SW}) as
a subleading correction in the large $L$ expansion. This can be done by
going back to the picture of loaded string (LS) and rederiving the mass
relation using expansion analysis of the LS model as in \cite{SelemW06}, but
with a different small parameter. The result is \cite{BChen:A15} (Appendix A)
\begin{equation}
(E-M_{Q})^{2}=\pi TL+a_{0},  \label{UReg}
\end{equation}%
with $a_{0}=(m_{l}+M_{Q}v_{Q}^{2})^{2}$ depending upon $v_{Q}\,$and two
effective masses of quarks, defined by
\begin{equation}
M_{Q}=\frac{m_{Q}}{\sqrt{1-v_{Q}^{2}}},m_{l}=\frac{m_{q}}{\sqrt{1-v_{q}^{2}}}%
,  \label{Eff}
\end{equation}%
($m_{q}$ is the bare mass of the light quark $u$ or $d,s$)

The relation (\ref{UReg}) was derived in Refs. \cite{BChen:A15,CWZ:C09}
for the singly heavy baryons. A phenomenological study \cite{KChenL:C18}
of the relation (\ref{UReg}) was made more recently for both heavy-light
mesons and singly heavy baryons, favoring (\ref{UReg}) to be a mass relation
mapping experimental data for all heavy-light hadron systems. When $L$ is very
large, Eq. (\ref{UReg})$\,$becomes $E-M_{Q}\simeq \sqrt{\alpha L/2}+a_{0}/%
\sqrt{2\alpha L}$, quite similar to the Selem-Wilczek relation (\ref{SW}),
whereas it avoids the singularity when $L\rightarrow 0$. For similar
discussions for the singly heavy baryons see Ref. \cite{BChen:A15}.
Considering the loaded string model is classical and valid only in the
quasi-classical region (with large quantum number), it is expected that (\ref%
{UReg}) needs quantum correction far from the quasi-classical region (e.g., $%
L\rightarrow 0$). However, the full quantum treatment is nontrivial for the
low-lying hadrons.

While the quasi-classical model can not give the intercept $\alpha (0)$ or $%
a_{0}$ of the trajectory, which was rooted in a quantum effect of the HL
system, the fact that Eqs. (\ref{MnJ}) and (\ref{HRegg}) with constant
intercept $a_{0}$($c$) maps the excited mass of light mesons quite well is
very impressing \cite{JohnsonN:PRD79,EFG:C10,SonnenscheinW:JHEP14}. Therefore,
it is hoped that a modified version of the quasi-classical prediction may fit
the mass spectrum in full quantum theory, if appropriately using the
Regge phenomenology of hadrons \cite{JohnsonN:PRD79,SonnenscheinW:JHEP14}.
In fact, in the string picture, the full quantum treatment provides merely
the "quantum defect" $\alpha (0)$ in $L$ \cite{Rebbi:PR74}.

Our approach for meson excitations is to find a Regge-like mass relation in
the quasi-classical region (at large-$L$ or $n$) at first, and then
extrapolate it to the lower-$L$ region by taking into account the
enhancement of the binding effect between the heavy-quark and strange quark
appropriately. This extrapolation relies on the empirical fact that the
hadronic trajectories are nearly linear even for lower excitations, though
the trajectory parameters may be flavor-dependent weakly.

To account for the measured mass data of the heavy-light mesons comprehensively,
it is necessary to consider the radial excitations. The simplest way is to assume
the Regge-like relation (\ref{MnJ}). Until now, no agreement is achieved for
the slope ratio $\beta /\alpha $. In Table \ref{rf}, we collect the predictions
or estimates of this ratio factor in the existing literatures quoted.

\renewcommand{\arraystretch}{1.8}
\begin{table}[!htbp]
\caption{The predictions or estimates of the ratio factor $%
\beta /\alpha $ in the existing literatures quoted.}\label{rf}
\begin{tabular}{p{1.0cm}p{1.8cm}p{1.0cm}p{4.3cm}}
\toprule[1pt]
Refs. & System & $\beta /\alpha $ & Method (framework) \\ \hline
\cite{Lovelace:B68} & Light-light & $1$ & Old dual amplitudes \\
\cite{Shapiro:69} & Light-light & $1$ & Old dual amplitudes \\
\cite{Veneziano:68} & Light-light & $1$ & Old dual amplitudes \\
\cite{AllenGO:D01} & Light-light & $2$ & Semiclassical quantization (String)
\\
\cite{Afonin:A07,Afonin:C07} & Light-light & $1$ & Phenomenological analysis (Exp.) \\
\cite{Bicudo:D07} & Light-light & $\pi /2$ & Semiclassical
Analysis (Potential QM) \\
\cite{Olsson:D97} & Heavy-light & $\sqrt{2}$-$\sqrt{6}$ & Semiclassical
Analysis (Potential QM) \\
\cite{Anisovich:D00} & Light-light & $\simeq 1$ & Linear fit (Exp.) \\
\cite{Masjuan:D12} & Light-light & $\simeq 1.23$ & Linear fit (Exp.) \\
\cite{Forkel:07a,Afonin:A17} & Light-light & $1$ & Holographic QCD \\
\cite{EFG:D09} & Light-meson & $1.3$ & Relativistic quasipotential (QM) \\
\cite{EFG:C10} & Heavy-light & $1.4$ & Relativistic quasipotential (QM). \\
\cite{EFG:D11} & Heavy baryon & $1.5$ & Relativistic
quasipotential (quark-diquark). \\
\cite{BChen:A15} & Heavy baryon & $1.57$ & The $v_{Q}$-expansion
(Relativistic String) \\
\bottomrule[1pt]
\end{tabular}
\end{table}

\medskip

The predictions of the ratio $\beta /\alpha $ are about $1$ mostly in the
light-light mesons while it is close to $1.5$ in the heavy-light mesons in
the literatures cited. In the case of heavy mesons, the actual value of
the ratio relies on how much the hadron mass is shifted, as the mass squared
$M^{2}$, instead of the shifted mass squared $(M-M_{Q})^{2}$, is nonlinear
in $L$, with slightly larger slope in the Regge plot near the low-$L$
excitations. Given the limited number of observed states, it of interest to
explore the flavor-dependence of the trajectory parameters in the heavy meson case.
The feature of our proposal (\ref{GRegg}) lies in:

(i) The flavor-dependence of the mass relation, which is embodied in the
effective mass of quarks, is incorporated to the mass-shifted Regge-like relation.
The correction to the mass relation (\ref{UReg}) due to the replacement (\ref{shift})
is tiny for the nonstrange $\bar{c}n$ and $\bar{b}n$ mesons for which
$m_{l}/M_{Q}$ is small. This is not the case for the strange $\bar{c}s$
and $\bar{b}s$ mesons. Qualitatively, it accounts for phenomenological energies
of heavy quark's binding with the strange quarks \cite{KarlinerR:D14,KarlinerR:D18},
approximately linear in the effective reduced mass $\tilde{\mu}_{lQ}$ of
the heavy quark and antiquark \cite{KarlinerR:18}.

(ii) The slope ponderation ($\pi :2$) between radial and angular trajectories
is examined successfully (Section \ref{sec3}) and enables us to accommodate
the newly-observed $D_{J}(3000)$ and $D_{J}^{\ast}(3000)$.

(iii) A prefactor $k$ for the light quark mass $m_{l}$ is added phenomenologically when the excited states of the HL mesons are involved. It employs to describe the dressing effect reduction of the light quark when it is highly excited.

\section{Numerical plots and test of the mass relations}\label{sec3}

To confront the mass relation (\ref{GRegg}) with the experimental
spectra of the HL mesons, we list, in Table \ref{dm} and \ref{bm},
the all observed charmed and charmed strange mesons and their masses,
and the corresponding predictions by the relativistic quasipotential
model \cite{EFG:C10}. In this work, some mesons, though they are listed,
do not enter during computing spin-averaged masses as their mass may be
shifted due to the near-threshold effects \cite{ChenCL:Rpp17}.
For example, the mesons $D_{s0}^{\ast }(2317)$ and $D_{s1}(2460)$ are not used
when spin-averaging of the $1P$ $D_{s}$ meson masses as they are light abnormally
in the light of the quark models, due probably to the near-threshold effects
or the exotic natures of these mesons (see \cite{ChenCL:Rpp17}, for recent review).

With the data in Table \ref{dm} and Table \ref{bm}, one can obtain
the spin-averaged (spin-AV.) masses by
\begin{equation}
\overline{M}_{nL}=\frac{1}{N_{nL}}\sum (2J+1)M_{nL}^{Exp}(J)  \label{Mex}
\end{equation}%
and then map the relation (\ref{GRegg}) of them by using the $\chi ^{2}$ fitting,
\begin{equation}
\chi ^{2}=\frac{1}{N}\sum \left( M_{nL}-\overline{M}_{nL}\right) ^{2},
\label{ch2}
\end{equation}%
in which the estimated mass, from the proposal (\ref{GRegg}), is given explicitly by
\begin{align}
M_{nL}  &  =\frac{M_{Q}}{1+km_{l}/M_{Q}}\nonumber\\
&  +\sqrt{\pi b\left( L+\frac{\pi }{2}%
n\right) +\left( m_{l}+M_{Q}\left( 1-\frac{m_{bareQ}^{2}}{M_{Q}^{2}}\right)
\right) ^{2}}  \label{MnL}
\end{align}%

Here, $N=13$ corresponds to the available number of the spin-averaged data
chosen from the Table \ref{rf} and \ref{dm}.

\renewcommand\tabcolsep{0.48cm}
\renewcommand{\arraystretch}{1.8}
\begin{table*}[!htbp]
\caption{The observed masses (in MeV) of the charmed and charmed strange
mesons \cite{Tanabashi:D18}. The some of quantum numbers indicated by
question marks are quark model predictions, which has not been established
experimentally. The errors less than $5$ MeV are not indicated.}\label{dm}
\begin{tabular}{ccccccc}
\toprule[1pt]
{\small State }$J^{P}$ & {\small Meson} & {\small Mass} &
{\small EFG \cite{EFG:C10}} & {\small Meson} & {\small Mass} &
{\small EFG \cite{EFG:C10}} \\
\hline
$%
\begin{array}{rr}
{\small 1}^{1}{\small S}_{0} & {\small 0}^{-} \\
{\small 1}^{3}{\small S}_{1} & {\small 1}^{-}%
\end{array}%
$ & $%
\begin{array}{r}
{\small D}^{\pm } \\
{\small D}^{\ast }{\small (2010)}^{\pm }%
\end{array}%
$ & $%
\begin{array}{r}
{\small 1869.7} \\
{\small 2010.3}%
\end{array}%
$ & $%
\begin{array}{r}
{\small 1871} \\
{\small 2010}%
\end{array}%
$ & $%
\begin{array}{r}
{\small D}_{s} \\
{\small D}_{s}^{\ast }{\small [J}^{P}?^{?}{\small ]}%
\end{array}%
$ & $%
\begin{array}{r}
{\small 1968.3} \\
{\small 2112.2}%
\end{array}%
$ & $
\begin{array}{r}
{\small 1969} \\
{\small 2111}%
\end{array}%
$ \\ $%
\begin{array}{rr}
{\small 1}^{3}{\small P}_{0} & {\small 0}^{+} \\
{\small 1P}_{1} & {\small 1}^{+} \\
{\small 1P}_{1} & {\small 1}^{+} \\
{\small 1}^{3}{\small P}_{2} & {\small 2}^{+}%
\end{array}%
$ & $%
\begin{array}{r}
{\small D}_{0}^{\ast }{\small (2400)}^{\pm } \\
{\small D}_{1}{\small (2430)}^{0} \\
{\small D}_{1}{\small (2420)}^{\pm }{\small [J}^{P}?^{?}{\small ]} \\
{\small D}_{2}^{\ast }{\small (2460)}^{\pm }%
\end{array}%
$ & $%
\begin{array}{r}
{\small 2351(7)} \\
{\small 2427(40)} \\
{\small 2423.2} \\
{\small 2465.4}%
\end{array}%
$ & $%
\begin{array}{r}
{\small 2406} \\
{\small 2469} \\
{\small 2426} \\
{\small 2460}%
\end{array}%
$ & $%
\begin{array}{r}
{\small D}_{s0}^{\ast }{\small (2317)} \\
{\small D}_{s1}{\small (2460)} \\
{\small D}_{s1}{\small (2536)} \\
{\small D}_{s2}^{\ast }{\small (2573)}%
\end{array}%
$ & $%
\begin{array}{r}
{\small 2317.7} \\
{\small 2459.5} \\
{\small 2535.1} \\
{\small 2569.1}%
\end{array}%
$ & $%
\begin{array}{r}
{\small 2509} \\
{\small 2574} \\
{\small 2536} \\
{\small 2571}%
\end{array}%
$ \\ $%
\begin{array}{rc}
{\small 2}^{1}{\small S}_{0} & {\small 0}^{-} \\
{\small 2}^{3}{\small S}_{1} & {\small 1}^{-}%
\end{array}%
$ & $%
\begin{array}{r}
{\small D(2550)}^{0}{\small [J}^{P}?^{?}{\small ]} \\
{\small D}^{\ast }{\small (2640)}^{\pm }{\small [J}^{P}?^{?}{\small ]}%
\end{array}%
$ & $%
\begin{array}{r}
{\small 2564(20)} \\
{\small 2637(6)}%
\end{array}%
$ & $%
\begin{array}{r}
{\small 2581} \\
{\small 2632}%
\end{array}%
$ & $%
\begin{array}{r}
\\
{\small D}_{s1}^{\ast }{\small (2700)}%
\end{array}%
$ & $%
\begin{array}{r}
\\
{\small 2708.3}%
\end{array}%
$ & $%
\begin{array}{r}
{\small 2688} \\
{\small 2731}%
\end{array}%
$ \\ $%
\begin{array}{rc}
{\small 1}^{3}{\small D}_{1} & {\small 1}^{-} \\
{\small 1D}_{2} & {\small 2}^{-} \\
{\small 1D}_{2} & {\small 2}^{-} \\
{\small 1}^{3}{\small D}_{3} & {\small 3}^{-}%
\end{array}%
$ & $%
\begin{array}{c}
\\
\\
{\small D(2740)}^{0}{\small [J}^{P}?^{?}{\small ]} \\
{\small D}_{3}^{\ast }{\small (2750)}%
\end{array}%
$ & $%
\begin{array}{c}
\\
\\
{\small 2737(12)} \\
{\small 2763.5}%
\end{array}%
$ & $%
\begin{array}{c}
{\small 2788} \\
{\small 2850} \\
{\small 2806} \\
{\small 2863}%
\end{array}%
$ & $%
\begin{array}{c}
{\small D}_{s1}^{\ast }{\small (2860)} \\
\\
\\
{\small D}_{s3}^{\ast }{\small (2860)}%
\end{array}%
$ & $%
\begin{array}{c}
{\small 2859(27)} \\
\\
\\
{\small 2860(7)}%
\end{array}%
$ & $%
\begin{array}{c}
{\small 2913} \\
{\small 2961} \\
{\small 2931} \\
{\small 2971}%
\end{array}%
$ \\ $%
\begin{array}{rr}
{\small 2}^{3}{\small P}_{0} & {\small 0}^{+} \\
{\small 2P}_{1}^{{}} & {\small 1}^{+} \\
{\small 2P}_{1} & {\small 1}^{+} \\
{\small 2}^{3}{\small P}_{2} & {\small 2}^{+}%
\end{array}%
$ & $
$ & $
$ & $
\begin{array}{c}
{\small 2919} \\
{\small 3021} \\
{\small 2932} \\
{\small 3012}%
\end{array}%
$ & $
\begin{array}{c}
\\
\\
{\small D}_{sJ}{\small (3040)}{\small [J}^{P}?^{?}{\small ]} \\
\\%
\end{array}%
$ & $
\begin{array}{c}
\\
\\
{\small 3044}_{-9}^{+31} \\
\\%
\end{array}%
$ & $
\begin{array}{c}
{\small 3054} \\
{\small 3154} \\
{\small 3067} \\
{\small 3142}%
\end{array}%
$ \\ $%
\begin{array}{rr}
{\small 3}^{3}{\small P}_{0} & {\small 0}^{+} \\
{\small 3P}_{1}^{{}} & {\small 1}^{+} \\
{\small 3P}_{1} & {\small 1}^{+} \\
{\small 3}^{3}{\small P}_{2} & {\small 2}^{+}%
\end{array}%
$ & $
\begin{array}{c}
\\
\\
\\
{\small D(3000)}^{0}{\small [J}^{P}?^{?}{\small ]}%
\end{array}%
$ & $
\begin{array}{c}
\\
\\
\\
{\small 3214(60)}%
\end{array}%
$ & $
\begin{array}{c}
{\small 3346} \\
{\small 3461} \\
{\small 3365} \\
{\small 3407}%
\end{array}%
$ & $
$ & $
$ & $
\begin{array}{c}
{\small 3513} \\
{\small 3618} \\
{\small 3519} \\
{\small 3580}%
\end{array}%
$ \\
\bottomrule[1pt]
\end{tabular}
\end{table*}

\medskip

\renewcommand\tabcolsep{0.5cm}
\renewcommand{\arraystretch}{1.8}
\begin{table*}[!htbp]
\caption{The observed masses of the bottomed and bottomed strange mesons \cite{Tanabashi:D18}.
The some of quantum numbers as shown are quark model
predictions. The mass is in MeV. The errors less than $5$ MeV are not indicated.}\label{bm}
\begin{tabular}{ccccccc}
\toprule[1pt]
{\small State }$J^{P}$ & {\small Meson} & {\small Mass} &
{\small EFG \cite{EFG:C10}} & {\small Meson} & {\small Mass} &
{\small EFG \cite{EFG:C10}} \\ \hline
$%
\begin{array}{rr}
{\small 1}^{1}{\small S}_{0} & {\small 0}^{-} \\
{\small 1}^{3}{\small S}_{1} & {\small 1}^{-}%
\end{array}%
$ & $%
\begin{array}{r}
B^{0} \\
B^{\ast }%
\end{array}%
$ & $%
\begin{array}{r}
{\small 5279.6} \\
{\small 5324.7}%
\end{array}%
$ & $
\begin{array}{r}
{\small 5280} \\
{\small 5326}%
\end{array}%
$ & $%
\begin{array}{r}
{\small B}_{s} \\
{\small B}_{s}^{\ast }%
\end{array}%
$ & $%
\begin{array}{r}
{\small 5366.9} \\
{\small 5415.4}%
\end{array}%
$ & $
\begin{array}{r}
{\small 5372} \\
{\small 5414}%
\end{array}%
$ \\ $%
\begin{array}{rr}
{\small 1}^{3}{\small P}_{0} & {\small 0}^{+} \\
{\small 1P}_{1} & {\small 1}^{+} \\
{\small 1P}_{1} & {\small 1}^{+} \\
{\small 1}^{3}{\small P}_{2} & {\small 2}^{+}%
\end{array}%
$ & $%
\begin{array}{r}
{\small B}_{J}^{\ast }{\small (5732)}{\small [J}^{P}?^{?}{\small ]} \\
\\
{\small B}_{1}{\small (5721)}^{0} \\
{\small B}_{2}^{\ast }{\small (5747)}^{0}%
\end{array}%
$ & $%
\begin{array}{r}
{\small 5698(8)} \\
\\
{\small 5726.0} \\
{\small 5739.5}%
\end{array}%
$ & $
\begin{array}{r}
{\small 5749} \\
{\small 5774} \\
{\small 5723} \\
{\small 5741}%
\end{array}%
$ & $%
\begin{array}{r}
\\
{\small B}_{sJ}^{\ast }{\small (5850)}{\small [J}^{P}?^{?}{\small ]} \\
{\small B}_{s1}{\small (5830)} \\
{\small B}_{s2}^{\ast }{\small (5840)}%
\end{array}%
$ & $%
\begin{array}{r}
\\
{\small 5853(15)} \\
{\small 5828.6} \\
{\small 5839.9}%
\end{array}%
$ & $
\begin{array}{r}
{\small 5833} \\
{\small 5865} \\
{\small 5831} \\
{\small 5842}%
\end{array}%
$ \\ $%
\begin{array}{rr}
{\small 2}^{1}{\small S}_{0} & {\small 0}^{-} \\
{\small 2}^{3}{\small S}_{1} & {\small 1}^{-}%
\end{array}%
$ & $%
\begin{array}{r}
{\small B}_{J}{\small (5840)}^{0}{\small [J}^{P}?^{?}{\small ]} \\
{\small B}_{J}{\small (5970)}^{0}{\small [J}^{P}?^{?}{\small ]}%
\end{array}%
$ & $%
\begin{array}{r}
{\small 5863(9)} \\
{\small 5971(5)}%
\end{array}%
$ & $
\begin{array}{r}
{\small 5890} \\
{\small 5906}%
\end{array}%
$ & $
$ & $
$ & $
\begin{array}{r}
{\small 5976} \\
{\small 5992}%
\end{array}%
$ \\
\bottomrule[1pt]
\end{tabular}
\end{table*}

\medskip

\renewcommand\tabcolsep{0.425cm}
\renewcommand{\arraystretch}{1.8}
\begin{table*}[!htbp]
\caption{The effective masses of quarks determined via mapping the Table \ref{rf}
and \ref{dm}. The mass is in GeV and $b$ is in GeV$^{2}$; $k=0.551$%
.}\label{em}
\begin{tabular}{cccccccccc}
\toprule[1pt]
{\small Parameters} & ${\small M}_{c}$ & ${\small M}_{b}$ & ${\small m}_{n}$
& ${\small m}_{s}$ & ${\small b}${\small (}${\small c\bar{n}}${\small )} & $%
{\small b}${\small (}${\small c\bar{s}}${\small )} & ${\small b(b\bar{n})}$
& ${\small b(b\bar{s})}$ & ${\small \chi }_{SM}$ \\ \hline
{\small This work} & ${\small 1.46}$ & ${\small 4.52}$ & ${\small 0.31}$ & $%
{\small 0.49}$ & ${\small 0.264}$ & ${\small 0.314}$ & ${\small 0.306}$ & $%
{\small 0.372}$ & ${\small 0.0116}$ \\
{\small EFG \cite{EFG:C10}} & ${\small 1.55}$ & ${\small 4.88}$ & ${\small 0.33}$ & $%
{\small 0.5}$ & ${\small 0.64}${\small /}${\small 0.58}$ & ${\small 0.68/0.64%
}$ & ${\small 1.25/1.21}$ & ${\small 1.28/1.23}$ &  \\
\bottomrule[1pt]
\end{tabular}
\end{table*}

\medskip

\renewcommand\tabcolsep{0.55cm}
\renewcommand{\arraystretch}{1.8}
\begin{table*}[!htbp]
\caption{The trajectory parameters ($\alpha ^{\prime },\alpha _{0}$) in (\ref%
{GRegg}) and predicted by Ref. \cite{EFG:C10}. The unit of the $\alpha
^{\prime }$ is in GeV$^{-2}$.}\label{tp}
\begin{tabular}{ccccc}
\toprule[1pt]
{\small Traj. Parameters} & $c\bar{n}${\small (natural J}$^{P}${\small )} & $%
c\bar{s}${\small (natural J}$^{P}${\small )} & $b\bar{n}${\small (natural J}$%
^{P}${\small )} & $b\bar{s}${\small (natural J}$^{P}${\small )} \\ \hline
{\small This work(}$\alpha ^{\prime },\alpha _{0}${\small )} & ${\small %
(1.21,-0.52)}$ & ${\small (1.01,-0.72)}$ & ${\small (1.04,-0.97)}$ & $%
{\small (0.86,-1.13)}$ \\
{\small EFG (}$\alpha ^{\prime },\alpha _{0}${\small ) \cite{EFG:C10}} & $%
\begin{array}{r}
{\small (0.494,-1.00(4))} \\
{\small (0.548,-3.21(12))}%
\end{array}%
$ & $%
\begin{array}{r}
{\small (0.469,-1.10(4))} \\
{\small (0.497,-3.16(12))}%
\end{array}%
$ & $%
\begin{array}{r}
{\small (0.254,-6.30(36))} \\
{\small (0.263,-8.77(47))}%
\end{array}%
$ & $%
\begin{array}{r}
{\small (0.249,-6.43(51))} \\
{\small (0.259,-8.87(58))}%
\end{array}%
$ \\
\bottomrule[1pt]
\end{tabular}
\end{table*}

\medskip

\renewcommand\tabcolsep{0.34cm}
\renewcommand{\arraystretch}{1.8}
\begin{table*}[!htbp]
\caption{The masses of the charmed and charmed strange mesons \cite%
{Tanabashi:D18}. The mass is in MeV.}\label{md}
\begin{tabular}{ccccccccc}
\toprule[1pt]
$J^{P}$ & Meson & Mass & CDLLM \cite{KChenL:C18} & Exp. (Spin-AV.) &
Meson & Mass & CDLLM \cite{KChenL:C18} & Exp. (Spin-AV.) \\ \hline
$(0,1)^{-}$ & $D(1S)$ & $1964$ & $1910.6$ & $1975.1$ & $D_{s}(1S)$ & $2071$ & $1991.6$ & $2076.2$
\\
$(0,1,2)^{+}$ & $D(1P)$ & $2430$ & $2441.6$ & $2436(12)$ & $D_{s}(1P)$ & $2532$ & $2516.4$ & $2556.4$
\\
$(0,1)^{-}$ & $D(2S)$ & $2624$ &  & $2619(10)$ & $D_{s}(2S)$ & $2733$ &  & $2708$
\\
$(1,2,3)^{-}$ & $D(1D)$ & $2752$ & $2758.3$ & $2752(7)$ & $D_{s}(1D)$ & $2868$ & $2843.5$ & $2860(13)$
\\
$(0,1,2)^{+}$ & $D(2P)$ & $2908$ &  &  & $D_{s}(2P)$ & $3032$ &  & $3044_{-9}^{+31}$ \\
$(0,1)^{-}$ & $D(3S)$ & $3049$ &  &  & $D_{s}(3S)$ & $3182$ &  & \\
$(2,3,4)^{+}$ & $D(1F)$ & $3015$ & $3007.7$ &  & $D_{s}(1F)$ & $3146$ & $3103.1$ & \\
$(1,2,3)^{-}$ & $D(2D)$ & $3148$ &  &  & $D_{s}(2D)$ & $3288$ &  & \\
$(0,1,2)^{+}$ & $D(3P)$ & $3273$ &  & $3214(60)$ & $D_{s}(3P)$ & $3420$ &  & \\
\bottomrule[1pt]
\end{tabular}
\end{table*}

\medskip

\renewcommand\tabcolsep{0.34cm}
\renewcommand{\arraystretch}{1.8}
\begin{table*}[!htbp]
\caption{The masses of the bottomed and bottomed strange mesons \cite%
{Tanabashi:D18}. The mass is in MeV.}\label{mb}
\begin{tabular}{ccccccccc}
\toprule[1pt]
$J^{P}$ & Meson & Mass & CDLLM \cite{KChenL:C18} & Exp. (Spin-AV.) &
Meson & Mass & CDLLM \cite{KChenL:C18} & Exp. (Spin-AV.) \\ \hline
$(0,1)^{-}$ & $B(1S)$ & $5320$ & $5303.1$ & $5313.4$ & $B_{s}(1S)$ & $5412$ & $5390.9$ & $5403.3$
\\
$(0,1,2)^{+}$ & $B(1P)$ & $5730$ & $5732.6$ & $5730$ & $B_{s}(1P)$ & $5840$ & $5834.5$ & $5840$
\\
$(0,1)^{-}$ & $B(2S)$ & $5917$ &  & $5944(6)$ & $B_{s}(2S)$ & $6039$ &  &  \\
$(1,2,3)^{-}$ & $B(1D)$ & $6044$ & $6009.2$ &  & $B_{s}(1D)$ & $6175$ & $6129.4$ &  \\
$(0,1,2)^{+}$ & $B(2P)$ & $6199$ &  &  & $B_{s}(2P)$ & $6342$ &  &  \\
$(0,1)^{-}$ & $B(3S)$ & $6342$ &  &  & $B_{s}(3S)$ & $6497$ &  &  \\
$(2,3,4)^{+}$ & $B(1F)$ & $6308$ & $6230.2$ &  & $B_{s}(1F)$ & $6459$ & $6366.9$ &  \\
$(1,2,3)^{-}$ & $B(2D)$ & $6443$ &  &  & $B_{s}(2D)$ & $6606$ &  &  \\
$(0,1,2)^{+}$ & $B(3P)$ & $6571$ &  &  & $B_{s}(3P)$ & $6744$ &  &  \\
\bottomrule[1pt]
\end{tabular}
\end{table*}

\medskip

\begin{figure}[!htbp]
\begin{center}
\includegraphics[width=3.4in]%
{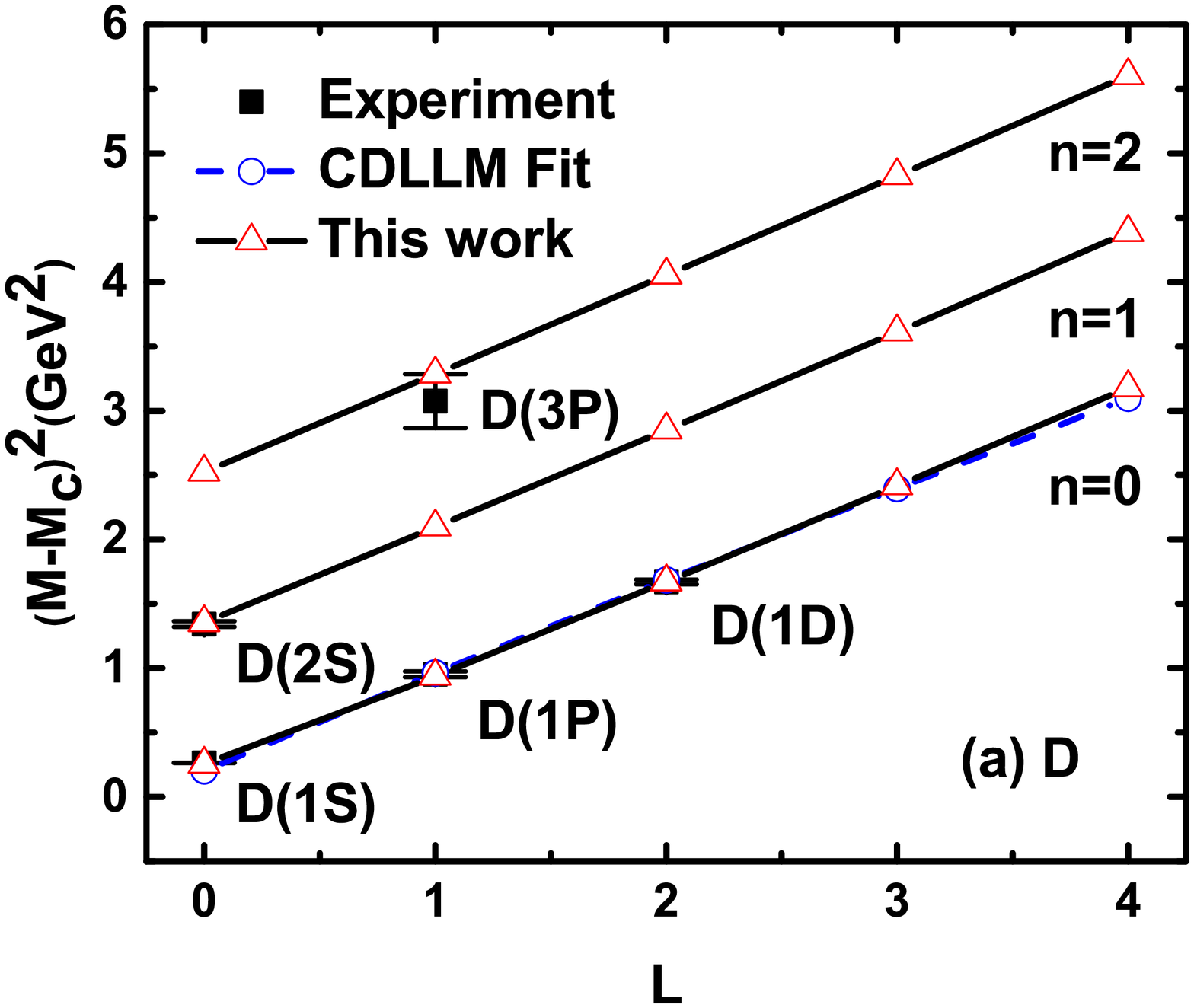}%
\caption{$D$ meson spectroscopy.}\label{FD}
\end{center}
\end{figure}

\medskip

\begin{figure}[!htbp]
\begin{center}
\includegraphics[width=3.4in]%
{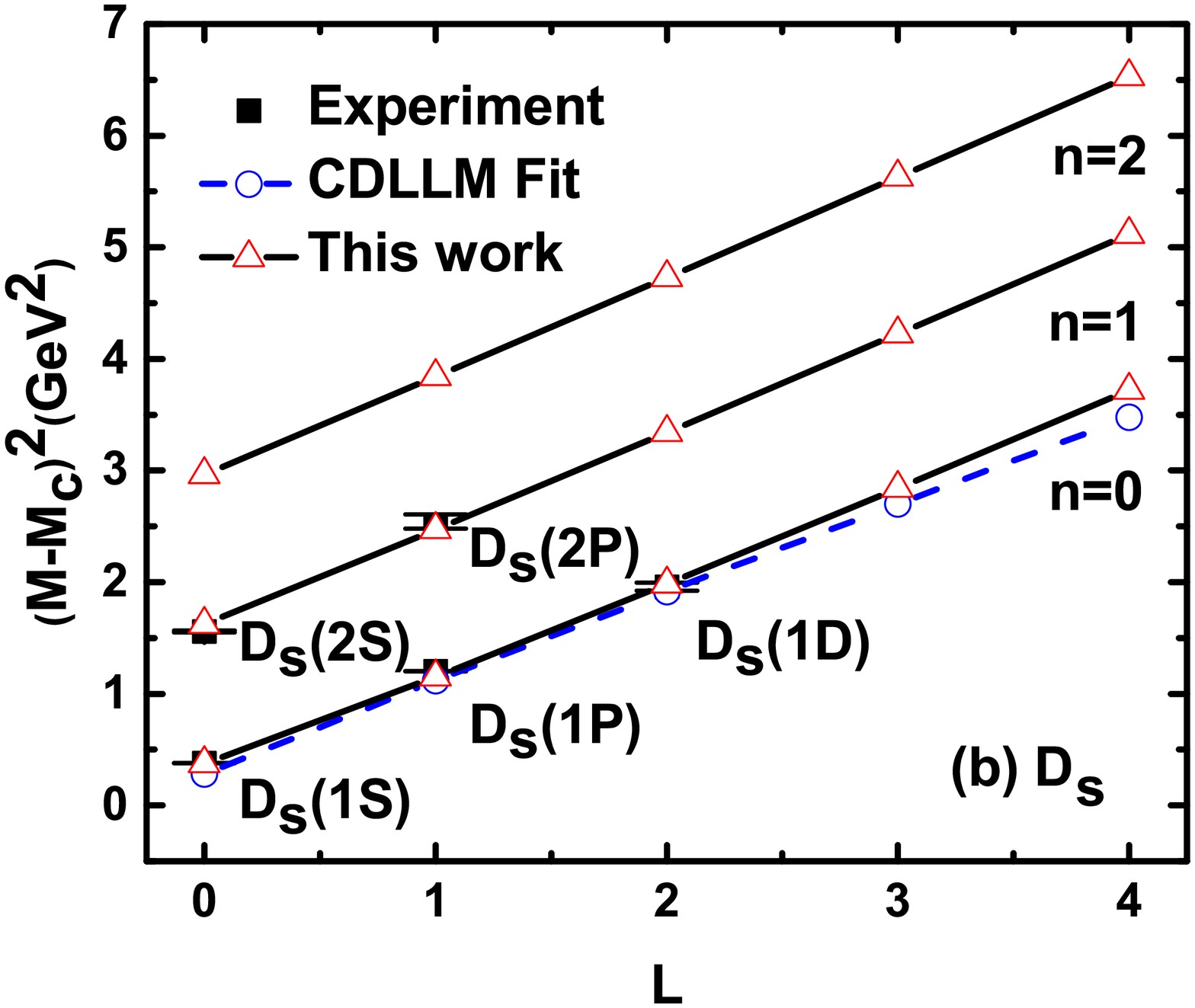}%
\caption{$D_{s}$ meson spectroscopy.}\label{FDs}
\end{center}
\end{figure}

\medskip

\begin{figure}[!htbp]
\begin{center}
\includegraphics[width=3.4in]%
{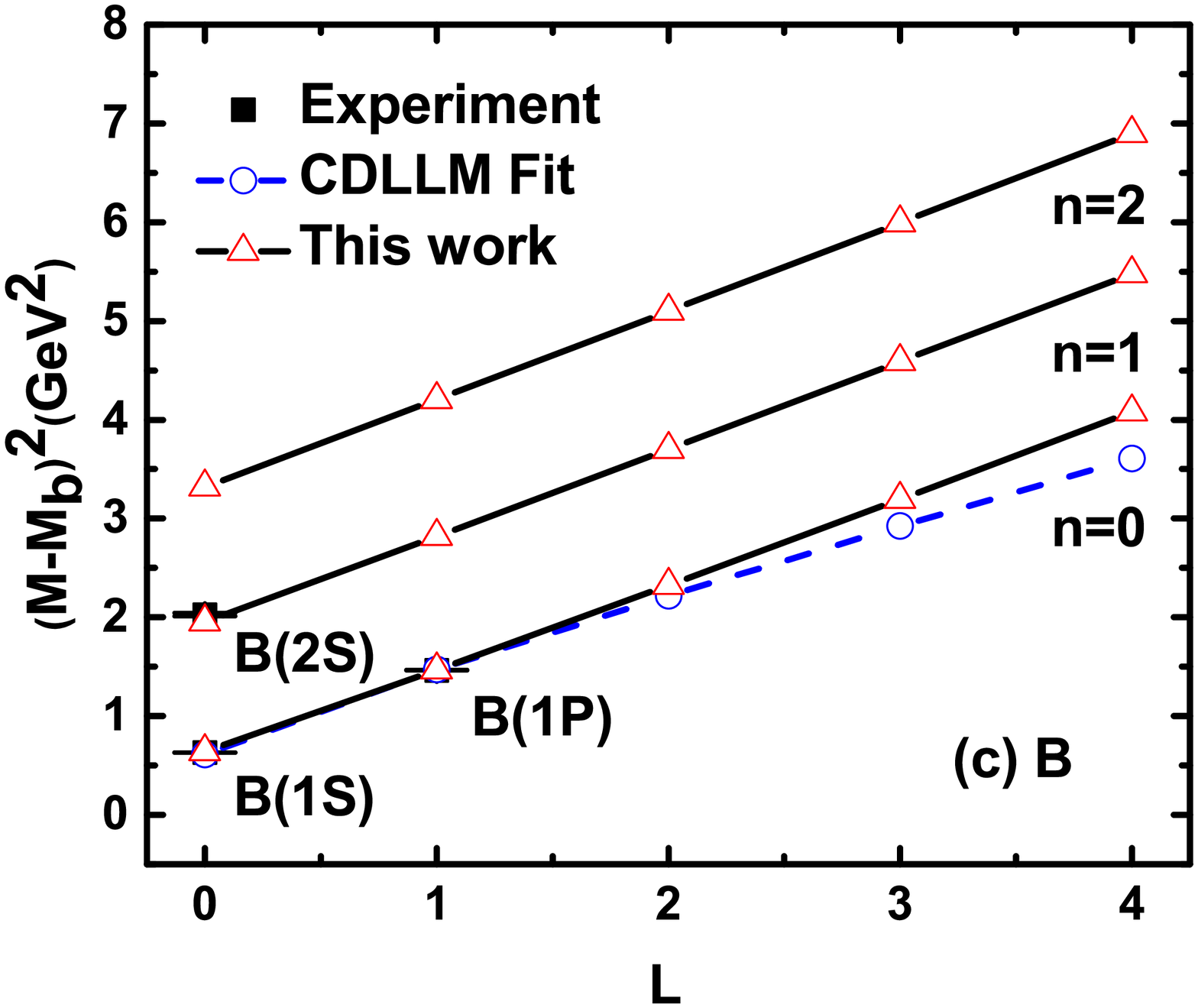}%
\caption{$B$ meson spectroscopy.}\label{FB}
\end{center}
\end{figure}

\medskip

\begin{figure}[!htbp]
\begin{center}
\includegraphics[width=3.4in]%
{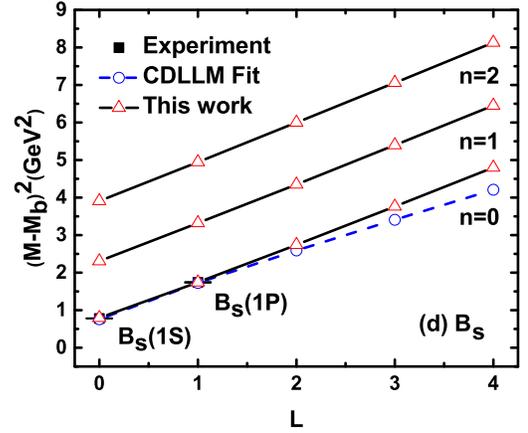}%
\caption{$B_{s}$ meson spectroscopy.}\label{FBs}
\end{center}
\end{figure}

\medskip

The mapped values of the parameters are listed in \ref{em} and that of
the ensuing trajectory parameters determined by the RHS of the relation
(\ref{GRegg}) are listed in Table \ref{tp}. The trajectory parameters
($\alpha ^{\prime },\alpha _{0}$) in Table \ref{tp} are related to
the parameters in Table \ref{em} through

\begin{equation}
\alpha ^{\prime }=\frac{1}{\pi b},\alpha _{0}=\frac{1}{\pi b}\left(
m_{q}+M_{Q}-m_{bareQ}^{2}/M_{Q}\right) ^{2}.  \label{alpha0}
\end{equation}

As is shown in Table \ref{em}, the mapped effective masses of the quarks in this
work agree remarkably with that in the quark model \cite{EFG:C10}. According to
the plots \cite{KChenL:C18} for the HL mesons, the Regge-like relation fitting of
the observed masses of the HL mesons by Eq. (\ref{UReg}) yields a close but
different trajectories (CDLLM) in contrast with that of Ref. \cite{EFG:C10}.
We show in Table \ref{md} the masses of the $D/D_{s}$ mesons determined by
Eq. (\ref{GRegg}) (denoted as this work) and by Eq. (\ref{UReg}) (denoted as CDLLM)
in Ref. \cite{KChenL:C18}, and the experimental masses of the spin-averaged,
and in Table \ref{mb} the corresponding masses of the $B/B_{s}$ mesons by
two equations and the experimental masses of the spin-averaged. The comparisons
are plotted in Figs. \ref{FD}-\ref{FDs} for the Table \ref{md} and in Figs.
\ref{FB}-\ref{FBs} for the Table \ref{mb}, correspondingly.

We end this section by giving the following remarks:

(1) As far as the spin-averaged spectra concern, the observed excited states
of the $D/D_{s}$ and $B/B_{s}$ mesons can be reasonably described by the
mass relation (\ref{MnL}), with the slope parameters depending weakly upon
the flavors.

(2) Once determined through the known data, the numerical values of the
parameters (the effective mass $m_{i}$($i=n,s,c,b$) of quarks and the slope
$b$) can be used to predict the spin-averaged masses of the higher
excitations of the HL mesons, as shown in Table \ref{md} and \ref{mb}.
For instance,the mass of the $D(2P)$ is about $2908$MeV and that of
the $D(2D)$ about $3148$MeV.

(3) The relation (\ref{GRegg}) gives a straightforward way to determine the
effective mass of quarks, which is nontrivial in the potential quark models.
The last term $a_{0}$ in Eq. (\ref{UReg}), when divided by $\pi T$, may
play the role of intercept of the trajectory considered, which embodies the
short-distance QCD correction to the QCD string model \cite{BakerS:D02}.

\section{The QCD string (flux tube) picture}\label{sec4}

We show in this section how the general mass relation (\ref{GRegg})
is related to the QCD string (flux tube) picture. For this, we rewrite the
relativistic Hamiltonian \cite{JohnsonN:PRD79,SelemW06} of the
quark-antiquark bound system as
\begin{equation}
H=\sqrt{\mathbf{p}^{2}+m_{Q}^{2}}+\sqrt{\mathbf{p}^{2}+m_{q}^{2}}+V^{string},
\label{HH}
\end{equation}%
\begin{equation}
V^{string}=\frac{T}{\omega }[\arcsin (\omega r_{q})+\arcsin (\omega r_{Q})]
\label{Hstring}
\end{equation}%
where $V^{string}$ is the relativistic energy of string with constant
tension $T$, tied to the heavy quark with mass $m_{Q}$ and light antiquark
with mass $m_{q}$. Here, $v_{i}\equiv \omega r_{i}$($i=Q,q$) denote the velocity
of the string end $i$ to which the quark $i$ is tied, and $\omega $ stands
for the angular velocity of the rotating system.

In the light of this work, we attribute the success of the LS picture in
\cite{SelemW06} to the fact that it appropriately encodes the nonperturbative
behavior of QCD \cite%
{Veneziano:68,Nambu:D74,Nambu:B79,VeseliO:B96,Afonin:A07,Afonin:C07} (See \cite%
{BChen:A15,SonnenscheinW:JHEP14,KChenL:C18} for recent study). Since states
with the large quantum numbers are semiclassical, in which the antiquark is
mostly at large distances from the quark, their energy levels obey the WKB
quantization condition or semiclassical mass relation (\ref{UReg}) with $L$
quantized. In the case of low-lying states, a modification to the
semiclassical predictions is required to incorporate the short-distance
correction to the string-like interquark interactions. For similar
discussion see Ref. \cite{LandauL:77} for the hydrogen-like atoms. We will
supply the LS picture with the short-distance binding effect between
heavy (anti)quark and strange quarks to improve the Regge-like mass relation
(\ref{UReg}) in the low-lying region.

When choosing $v_{Q}$, the velocity of the string end $\bar{Q}$, as the
expanding parameter, which is conserved in the heavy quark limit, one can
show in the LS picture (see Appendix A)
\begin{equation}
(E-M_{Q})^{2}=\pi \sigma L+\left[ m_{l}+M_{Q}\left( 1-\frac{m_{bareQ}^{2}}{%
M_{Q}^{2}}\right) \right] ^{2},  \label{ERhl}
\end{equation}%

This gives the relation (\ref{UReg}). Considering that this is a mass
relation both in the large limit of the quantum number and in the massless
limit of the light quark, one can incorporate the short-distance effect due
to the two-body binding interaction using the replacement (\ref{shift}), so
that
\begin{equation}
M_{nL}=M_{Q}-\tilde{\mu}_{lQ}+\sqrt{\pi bL+a_{0}}.  \label{RepM}
\end{equation}%

For the angularly excited HL mesons, (\ref{RepM}) gives the Regge-like
relation (\ref{GRegg}) (with $n=0$). For the radially excited HL mesons
($n>0$), we use the quasi-classical WKB approach for the string model Hamiltonian
(\ref{HH}).

Since $\arcsin (\omega r_{q})/\omega $ behaves linear in $r_{q}$ when the
velocity $v_{q}=\omega r_{q}$ of the light quark approaches the speed of
light: $v_{q}\simeq 1$, the potential (\ref{Hstring}) leads to an
asymptotically linear confining potential for large $r_{q}$. This is,
however, not true for the heavy quark part of potential since since $v_{Q}<1$
notably. Up to the leading order of Eq. (A5), one has for the heavy quark, %
$T/\omega =M_{Q}\omega r_{Q}$, which yields
\begin{equation*}
\omega =\sqrt{\frac{T}{M_{Q}r_{Q}}},v_{Q}=\sqrt{\frac{Tr_{Q}}{M_{Q}}},
\end{equation*}%
\begin{equation}
Tr_{Q}\frac{\arcsin (v_{Q})}{\omega }\simeq Tr_{Q}+\frac{(Tr_{Q})^{2}}{6M_{Q}%
}.  \label{TQ}
\end{equation}

For the string energy part of the light quark, one has
\begin{equation}
Tr_{q}\frac{\arcsin (v_{q})}{v_{q}}\simeq \frac{\pi }{2}Tr_{q}\text{, When }%
v_{q}\simeq 1.  \label{Tq}
\end{equation}%

It follows from (\ref{TQ}) and (\ref{Tq}) that the string energy for high-$L$
states becomes
\begin{equation}
V^{string}(r_{q},r_{Q})=\frac{\pi }{2}Tr_{q}+Tr_{Q}+\frac{(Tr_{Q})^{2}}{%
6M_{Q}}.  \label{Vqq}
\end{equation}%

Comparing (\ref{Vqq}) with the string energy for the light-light mesons,
\begin{equation*}
V^{string}(r_{q},r_{q})\simeq \frac{\pi }{2}Tr_{q}+\frac{\pi }{2}Tr_{q}=%
\frac{\pi }{2}Tr,
\end{equation*}%
one sees that the tensions for light quark $q$ and for the heavy antiquark
$\bar{Q}$ differ by a factor of $\pi /2$. Using $r_{q}=r-r_{Q}$, the total
Hamiltonian (\ref{HH}) of the bound system becomes, up to a $1/M_{Q}$
correction,
\begin{equation}
H=M_{Q}+\sqrt{\mathbf{p}^{2}+m_{q}^{2}}+\frac{\pi }{2}Tr+Tr_{Q}\left[ 1-%
\frac{\pi }{2}\right] .  \label{Hred}
\end{equation}

Given that the recoil effect has been taken into account in (\ref{shift}),
one can choose $r_{Q}\rightarrow 0$ in the heavy quark limit, so that
the problem becomes that of the light quark with Hamiltonian
\begin{equation}
H=M_{Q}+|\mathbf{p|}+\frac{\pi }{2}Tr,  \label{Hred2}
\end{equation}%
where the chiral limit($m_{q}\rightarrow 0$) has been taken. Assuming the
orbital angular momentum $L$ of a meson is dominated by that of the string,
which is used in the n\"{a}tive rotating string picture \cite%
{GoddardRB:B73,Nambu:D74,Goto:71},
\begin{equation*}
l_{q}=\frac{m_{q}v_{q}^{2}/\omega }{\sqrt{1-v_{q}^{2}}}=\omega
m_{l}r_{q}^{2}\ll L_{string},
\end{equation*}%
one can take $l_{q}\rightarrow 0$ so that $|\mathbf{p|}%
^{2}=p_{r}^{2}+l_{q}(l_{q}+1))/r^{2}\simeq p_{r}^{2}$. This reduces the
radial version of WKB approximation to that of one-dimensional effectively,
in which the CM of meson is located at $r_{Q}$ and the light quark $q$
moving in the force field of $\pi Tr/2$, with $r$ ranging from $0$ to $%
\infty $. The WKB quantization condition for (\ref{Hred2}) gives \cite%
{Arriola:JA07,BakerS:D02}
\begin{equation}
2\int_{0}^{(E-M_{Q})/a}(E-M_{Q}-a|x|)dx=\pi (n+b),  \label{WKB}
\end{equation}%
with $a\equiv \pi T/2$, that is,
\begin{equation}
(E-M_{Q})^{2}=\pi a\left( n+b\right) =\pi T\left( \frac{\pi }{2}n+\frac{\pi
}{2}b\right) ,  \label{WK}
\end{equation}%
where $r=(E-M_{Q})/a$ corresponds to the physically possible turning point.
This confirms the slope ratio $\pi /2$ in the relation (\ref{GRegg}) by
simply comparing (\ref{WK}) with the relation (\ref{UReg}) for the angular
excitations.

\section{Summary and discussions}\label{sec5}

We study the Regge-like spectra of singly heavy mesons, that is, the
$D/D_{s}$ and $B/B_{s}$ mesons, by proposing a general Regge-like mass
relations in which the slope ratio between the radial and angular-momentum
Regge trajectories is $\pi /2$ and the binding effect of heavy quark and
flavored light quarks has been taken into account. We test the the proposed
mass relation against the spin-averaged observed data of the singly heavy
mesons in their radially and angularly excited states and find that the
agreement is remarkable for the reasonable values of effective quark masses.
An argument is outlined for the mass relations using semiclassical WKB analysis of the relativistic interquark dynamics in the QCD string (flux tube) picture.

Some new predictions are made for more excited excitations. For instance,
the $D(3000)^{0}$ is more likely to be $3P$ state, and
the $B_{J}(5840)$ and $B_{J}(5970)$ can be the candidates of $2S$.
It is expected that the forthcoming Belle II and LHCb experiments can
test our predictions.

We note that the limitation of our mass relation (\ref{GRegg}) may stem from that it simply assumes the short-distance binding between quarks to be linear in the reduced quark mass, which is confirmed only for the ground S-wave states \cite{KarlinerR:D14,KarlinerR:D18}. To embody the possible deviations from the linearity for the highly-excited HL mesons, a simple prefactor $k$ for the light quark mass is employed which may vary with $L$ or $n$. We have checked that this dependence is weak in the case of the relation (\ref{GRegg}). Moreover, our predictions can not distinguish the spin multiplets due to neglecting of the spin-dependent interactions, as shown in Table VI and Table VII. The predictions by our relation (\ref{GRegg}), however, is generic in that they exempt the subtle influences due to the near-threshold effects hidden in the strange HL meson families(the $D_{s0}^{\ast }(2317)$ and $D_{s1}(2460)$ are avoided to use when  
when spin-averaging). We await the further explorations in the future study.


\section*{ACKNOWLEDGMENTS}

D. J thanks Bing Chen, X Liu and Atsushi Hosaka for many useful discussions and valuable comments. D. J is supported by the National Natural Science Foundation of China under the no. 11565023 and the Feitian Distinguished Professor Program of Gansu (2014-2016). W. D is supported by Undergraduate Innovative Ability Program 2018 (Grants No. CX2018B338).

\section*{APPENDIX A}

For the orbital excitions, the classical energy (\ref{HH}) and orbital
angular momentum for the loaded string can be written as \cite%
{JohnsonN:PRD79,SelemW06,LaCourseO:D99}
\begin{equation}
E=\frac{m_{Q}}{\sqrt{1-v_{Q}^{2}}}+\frac{m_{q}}{\sqrt{1-v_{q}^{2}}}+\frac{T}{%
\omega }[\arcsin (v_{q})+\arcsin (v_{Q})],  \tag{A1}
\end{equation}%
\begin{equation}
L=\sum_{i=Q,q}\frac{m_{i}v_{i}^{2}/\omega }{\sqrt{1-v_{i}^{2}}}+\frac{T}{%
\omega ^{2}}\sum_{i=Q,q}\int_{0}^{v_{i}}\frac{u^{2}du}{\sqrt{1-u^{2}}},
\tag{A2}
\end{equation}%
where $m_{Q,q}$ are the bare masses of the heavy and light quarks and $%
v_{i}=\omega r_{i}$($i=Q,q$). The Selem-Wilczek relation (\ref{SW}) follows
from \cite{SelemW06} upon the ($m_{i}\omega /T$)-expansion of (A1) and (A2).
Applying (\ref{Eff}), one has%
\begin{equation}
E=M_{Q}+m_{l}+\frac{T}{\omega }[\arcsin (v_{q})+\arcsin (v_{Q})],  \tag{A3}
\end{equation}%

The orbital angular momentum $L$ of the system is \cite%
{SelemW06,LaCourseO:D99}
\begin{equation}
L=\frac{1}{\omega }(M_{Q}v_{Q}^{2}+m_{l}v_{q}^{2})+\frac{T}{\omega ^{2}}%
\sum_{i=Q,q}\left[ \arcsin (v_{i})-v_{i}\sqrt{1-v_{i}^{2}}\right] ,  \tag{A4}
\end{equation}%
in which the last term is the orbital angular momentum due to the string
rotating.

The boundary condition of string at ends with heavy quark gives

\begin{equation}
\frac{T}{\omega }=\frac{m_{Q}v_{Q}}{1-v_{Q}^{2}},  \tag{A5}
\end{equation}%
which implies,
\begin{equation}
\frac{T}{\omega }=\frac{M_{Q}v_{Q}}{\sqrt{1-v_{Q}^{2}}}\simeq M_{Q}v_{Q}+%
\frac{1}{2}M_{Q}v_{Q}^{3}\text{.}  \tag{A6}
\end{equation}

Expanding Eqs. (A3) and (A4) up to $v_{Q}^{4}$,
\begin{align}
E &=&M_{Q}+m_{l}+\frac{\pi \sigma }{2\omega }+\frac{T}{\omega }\left[ v_{Q}-%
\frac{m_{q}}{m_{l}}+\frac{1}{6}v_{Q}^{3}\right] +\mathcal{O}[v_{Q}^{5}],
\tag{A7}\\
\omega L &=&m_{l}+M_{Q}v_{Q}^{2}+\frac{T}{\omega }\left[ \frac{\pi }{4}-%
\frac{m_{q}}{m_{l}}\right] +\frac{T}{3\omega }v_{Q}^{3}+\mathcal{O}%
[v_{Q}^{5}].  \tag{A8}
\end{align}%

With the help of Eq. (A6), Eqs. (A7) and (A8) become

\begin{eqnarray*}
E &=&M_{Q}+m_{l}+M_{Q}v_{Q}^{2}+\frac{\pi T}{2\omega }-\frac{m_{q}}{m_{l}}%
M_{Q}v_{Q}+\mathcal{O}[1/M_{Q}^{3}], \\
\omega L &=&m_{l}+M_{Q}v_{Q}^{2}+\frac{T}{\omega }\left( \frac{\pi }{4}-%
\frac{m_{q}}{m_{l}}\right) -\frac{m_{q}}{m_{l}}M_{Q}v_{Q}+\mathcal{O}%
[1/M_{Q}^{3}],
\end{eqnarray*}%
which, upon eliminating $\omega $ and ignoring the small term $m_{q}/m_{l}$,
leads to
\begin{equation}
(E-M_{Q})^{2}=\pi \sigma L+\left( m_{l}+\frac{P_{Q}^{2}}{M_{Q}}\right)
^{2}-2m_{q}P_{Q},  \tag{A9}
\end{equation}%
with $P_{Q}\equiv $ $M_{Q}v_{Q}=m_{l}v_{q}$ the conserved momentum of the
heavy quark relativistically. Rewriting the velocity $_{Q}^{2}=1-x_{Q}^{2}$
in terms of the mass ratio $x_{Q}=m_{Q}/M_{Q}$, Eq. (A9) yields (\ref{UReg})
or (\ref{ERhl}), where the bare mass term $2m_{q}P_{Q}$ has been ignored.

\end{document}